\RequirePackage{lineno}
\documentclass[twocolumn,prl,floatfix]{revtex4}
\usepackage{amsmath}
\usepackage{bm}
\usepackage{euscript}
\usepackage{graphicx}
\graphicspath{{./Figures/}}

\begin{document}


\title{Analysis of single-photon and
linear amplifier detectors for microwave cavity dark matter axion
searches}
\author{S.K. Lamoreaux}
\email{steve.lamoreaux@yale.edu}
\affiliation{Yale University, Department of Physics, P.O. Box 208120, New Haven CT 06520-8120, USA}
\author{K.A. van Bibber}
\affiliation{Department of Nuclear Engineering, University of
California Berkeley, Berkeley, CA 94720 USA}
\author{K.W.  Lehnert}
\affiliation{JILA and Department of Physics, University of Colorado and National Institute of Standards and Technology, Boulder, Colorado, 80309, USA  }
\author{G. Carosi}
\affiliation{Physical and Life Sciences Directorate, Lawrence
Livermore National Laboratory, Livermore, CA 94550 USA}

\begin{abstract}
We show that at higher frequencies, and thus higher axion masses,
single-photon detectors become competitive and ultimately favored,
when compared to quantum-limited linear amplifiers,
as the detector technology in microwave cavity experimental searches for galactic halo dark
matter axions. The cross-over point in this comparison is of order 10 GHz ($\sim 40\ \mu$eV), not
far above the frequencies of current searches.

\end{abstract}

\date{\today}

\maketitle

\section{Introduction}

The next generation of microwave cavity experimental searches for axions
constituting the dark matter halo of our galaxy will possess the
sensitivity to find or exclude favored axion models over a significant
fraction of their allowed mass ranges. These experiments will benefit
from new tunable cavity designs of higher frequency, and possibly much
higher quality factor $Q$ with the use of thin-film superconducting
coatings. They will also operate with much lower intrinsic noise owing
to the development of quantum-limited amplifiers, such as
Microstrip-coupled SQUID amplifiers (MSA), and Josephson Parametric
Amplifiers (JPA). The purpose of this short note is to demonstrate that
photon detectors will eventually win over linear amplifiers
at high frequencies, and possibly not far above where ADMX (Axion Dark Matter eXperiment) and ADMX-HF (High Frequency) will soon begin taking data ($\sim 1$ and $\sim 5$ GHz respectively).  Given that the microwave cavity experiments owe their extraordinary sensitivities to being both resonant and
spectrally-resolved, the possible utility of detectors which sacrifice
phase information, all or in part, bears some discussion.  Here we are concerned with
{\it fundamental} detection limits that cannot be improved upon, and we do not address
excess technical noise that can be eliminated with careful experimental design. We do note, however,
that technical noise problems tend to be more simply solved at higher frequencies.

In Sikivie's microwave cavity experiment, axions resonantly convert into
a very weak quasi-monochromatic microwave signal in a high-$Q$ cavity
permeated by a strong magnetic field (Figure 1) ~\cite{1}.
\begin{figure}[h!]
\includegraphics[width=\columnwidth]{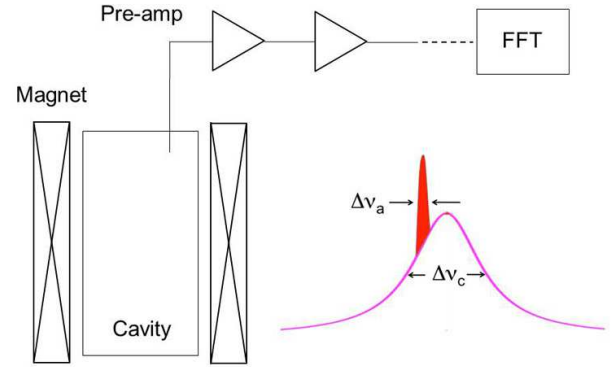}
\caption{Simplified schematic of the microwave cavity search for halo
axions. The insert depicts the virialized axion signal ($Q_a\equiv
E/\Delta E\sim 10^{6}$) within the Lorentzian bandpass
of the cavity ($Q_{c} = {Q}/(1+\beta)$).}
\end{figure}
The axion-photon conversion signal power is given by
\begin{linenomath}\begin{equation}
P_{a\rightarrow \gamma}= \eta g^2_{a\gamma\gamma}\left[{\rho_a\over m_a}\right]B_0^2 VCQ_c
\end{equation}\end{linenomath}
where $m_a$ and $\rho_a$ are the mass and
local density of the axion respectively, $g_{a\gamma\gamma}$ the
axion-photon coupling, $B_{0}$ is the
strength of the magnetic field, $V$ the cavity volume, and $C$ a
mode-dependent form factor. The loaded (coupled) quality factor of the
cavity is designated by $Q_c = Q/(1+\beta)$,\  $Q$ being the unloaded or intrinsic
quality factor, and $\beta$ the ratio of power coupled out by the
antenna to power dissipated by cavity wall losses. The fraction of the
converted power which is coupled out to the amplifier or other detector is
thus $\eta = \beta/(1+\beta)$.

The integration time required to achieve a desired signal-to-noise ($SNR$) ratio
is given by the Dicke radiometer equation, \cite{dicke}
\begin{linenomath}\begin{equation}
SNR={P_{a\rightarrow \gamma}\over k_BT}\sqrt{t\over \Delta \nu}
\end{equation}\end{linenomath}
where the system noise temperature $T$ is the sum of the physical
temperature plus the equivalent noise temperature of the device
$T = T_P + T_N$, $k_B$ is Boltzmann's constant, $t$ is the integration time, and
$\Delta \nu$ is the signal bandwidth. Here linear amplifiers and
single-photon detectors differ in two important respects. The first is the relevant
bandwidth determining the integration time, which is $\Delta \nu_a\sim \nu / Q_a$
in the case of a spectral receiver. A single-photon detector would measure the power over the
entire cavity bandpass, for which the relevant bandwidth is $\Delta \nu_c=\nu/Q_c$. Given that
in practice $Q_a / Q_c \sim 10-50$ even
for good copper cavities, a single-photon detector might not immediately suggest
itself. However, recent R\&D on the deposition of Type-II
superconducting thin films (e.g. Nb$_{x}$Ti$_{1-x}$N) on the
axial surfaces of microwave cavities, i.e. barrel and tuning rods, holds
out promise that hybrid normal-superconducting cavities would have a
bandpass approximately the width of the virialized axion signal,
$Q_a/ Q_c\sim 3$ ~\cite{2}.

More importantly, linear receivers are subject to quantum fluctuations,
leading to a standard quantum limit (SQL) in their performance. This limitation
is parameterized as an effective minimum temperature, $T_{SQL}=h\nu/k_B$ for any linear amplifier \cite{haus,caves}.  The origin of the SQL can be understood as follows.  Consider a well defined mode of an electromagnetic field.  In analogy with the harmonic oscillator and following \cite{caves}, we introduce conjugate variables $p_0$ and $q_0$ as operators that determine the field; the operators commute as $[p_0,q_0]=i\hbar/2$, which is true for any field.  Let us now apply a linear amplification to the zero point field, which increases each operator, hence the field, by a factor $G$, which motivates writing new field operators $p_f,\ q_f$ as
\begin{linenomath}\begin{equation}\label{noise}
[p_f,q_f]={i\hbar \over 2}= [Gp_0,Gq_0]+[p_g,q_g]={i G^2 \hbar \over 2}+[p_g,q_g]
\end{equation}\end{linenomath}where $[p_g,q_g]$ is introduced because linear amplification modifies the commutation relation. We see immediately that
\begin{linenomath}\begin{equation}
[p_g,q_g]={i(1-G^2)\hbar\over 2}.
\end{equation}\end{linenomath} By use of the generalized Heisenberg uncertainty relation, the bounds on the fluctuations on $p_g$ and $q_g$ are determined by the value of their commutator,
\begin{linenomath}\begin{equation}
\langle |\Delta p_g|^2\rangle\langle|\Delta q_g|^2\rangle \geq{(G^2-1)^2\hbar ^2\over 4}
\end{equation}\end{linenomath} (the zero point is a minimum uncertainty state, so the equality holds).
The first term on the right hand side of Eq. (\ref{noise}) results in a final field energy of $G^2 h\nu/2$, while the second term results in, by analogy, a field energy of $(G^2-1)h\nu/2$.  Referring the final field energy to the initial field energy before amplification,
\begin{linenomath}\begin{equation}
{1\over G^2}\left[{G^2 h \nu \over 2}+ {(G^2-1)h \nu \over 2}\right]={h\nu\over 2}+{(G^2-1)h\nu\over 2G^2}\approx 2\ \times\ {h\nu\over 2}
\end{equation}\end{linenomath}in the limit of large $G$. This result shows final state fluctuations that imply twice the zero point energy for the initial state; the is the SQL.  A principal implication is that the final state is, as might be expected, not a minimum uncertainty, or coherent, state.

In contrast to linear receivers,
single-photon detectors--which provide a strongly coupled measurement of a quantum nondemolition
variable--can in principle be arbitrarily noiseless. In the extreme case, $T\sim T_P \ll T_{SQL}$,
it is easily seen that the scanning rate
for single-photon detection relative to that for linear amplifiers improves exponentially with
frequency, by a factor of $\approx (Q_c/Q_a)\cdot \exp(2h\nu/k_BT_P)$
(in the low temperature limit) and will be overwhelmingly advantageous at the
high end of the open mass range. The main point of this note, however is
to examine quantum noise in the cavity and amplifier more rigorously,
and to conclude that single-photon detectors may be favorable even at
frequencies not far above the current search region, i.e. $< 10$ GHz.

\section{Detailed Analysis}

Here we present a detailed analysis of the ultimate sensitivity of two
schemes for the detection of electromagnetic field energy, and a
discussion of the conditions under which the sensitivity of one prevails
over the other. Numerical estimates will be drawn from the example of
the ADMX-HF  experiment nearing commissioning. This
platform is designed specifically to search for axions in the 20-100 $\mu$eV
range (5-25 GHz), as well as serving as a test-bed for new cavity and
detector concepts ~\cite{3}.

A linear detector provides a direct and coherent measurement of the EM field, and is thus subject to phase/amplitude zero point fluctuation noise. In contrast, a single-photon detector is sensitive only to the photon number, not the signal phase. Examples of single-photon detectors are photomultiplier tubes, and techniques that determine the photon number in a cavity (e.g., Rydberg atoms which respond to the square of the RF field amplitude). However, at non-zero temperatures, single-photon detectors are subject to noise from fluctuations in the number of detected thermal photons.  We further assume that the detected axion signal is the same for all detection techniques, so we need only compare the noise power for the different techniques. Finally, we assume that the noise temperature $T_N\ll T_P$ so the temperature $T$ is the same as the physical temperature of the system.

The detected noise power $P_\ell$ from a linear amplifier system is given by the
Dicke radiometer equation, which gives the fluctuations in the average noise power that is detected,
\begin{linenomath}\begin{equation}
P_\ell =k_BT\sqrt{\Delta \nu_a\over t}
\end{equation}\end{linenomath}
where the parameters have been defined in relation to Eq. (2).
For all axion cavity experiments to date, $Q_a$ associated with the axion is much larger than that of the cavity $Q_c$, typically for ADMX-HF, $Q_a\approx 10^6$ compared to the cavity $Q_c\approx 5\times 10^4$.  The axion $Q_a$ is determined by the virialized axion de Broglie wavelength (around 100 meters).  This means that the axion field will produce a coherent signal with a coherence time longer than the cavity lifetime; we think of the cavity in the applied magnetic field as a converter of axion field to a radiofrequency field, with the cavity being a passive intermediary.

In the case of low temperatures, the Dicke radiometer equation must be modified to account for zero point fluctuations. The photon occupation number of a cavity resonant at frequency $\nu$  at temperature $T$ is
\begin{linenomath}\begin{equation}
n(\nu,T) ={1\over e^{ h\nu /k_BT}-1}+{1\over 2}\equiv \bar n +{1\over 2}
\end{equation}\end{linenomath}
For a linear amplifier, the zero-point fluctuations contribute twice to the quantum-limited output noise~\cite{haus,caves}.
This motivates an ad hoc modification of the Dicke radiometer equation as
\begin{linenomath}\begin{equation}
P_\ell=h\nu(\bar n +1)\sqrt{\Delta \nu_a\over t}
\end{equation}\end{linenomath}
which produces the Standard Quantum Limit (SQL) of a linear amplifier system, and produces the high temperature limit given in Eq. (3).

For an experiment tuned to 5 GHz (e.g., ADMX-HF), and  operating at a realistic physical temperature of 20 mK, assuming $T_N\approx 0$ for an amplifier operating at the SQL,
$\bar n \approx  6\times 10^{-6}$, so the total effective field noise corresponds to $\bar n +1$ photons.
However, for photon-counting experiments where the cavity photon number is detected, the zero point fluctuations that lead to the SQL do not contribute noise; photon-counting experiments are subject primarily to shot noise on the detected thermal photons.  Given a detection efficiency $\eta$ that depends on the detection method and is of order unity, the rate of photon detection is
\begin{linenomath}\begin{equation}
\dot n=\eta \Gamma \bar n
\end{equation}\end{linenomath}
where $\Gamma=1/\tau_c$ is the loaded cavity lifetime.
(Here it is assumed that there is  no excess noise {\it i.e.}, there are no fluctuations in detection efficiency, no ``dark" noise with single photon detection, and that the excess noise temperature is zero.)

In an observation time $t$, $N$ photons (both associated with the axion and with the thermal excitations) are detected, where
\begin{linenomath}\begin{equation}
N=\dot n t=\eta \bar n\Gamma t
\end{equation}\end{linenomath}
which has fluctuations due to counting statistics of
\begin{linenomath}\begin{equation}
\delta N=\sqrt{N}= \sqrt{\eta \bar n\Gamma t}.
\end{equation}\end{linenomath}
Note that even though $\bar n$ is small, $N$ is almost always large enough to use Poisson statistics: again taking 5 GHz, 20 mK, and $t\approx 100$ sec, $\Gamma\approx 10^5$ sec$^{-1}$, and $\bar n\approx 6\times 10^{-6}$, $N\approx 60$ which is sufficiently large to use the $\sqrt{n}$ approximation for Poisson statistics.

As an aside, and for comparison, the detected axion signal itself will have shot noise, for example if $P_{a\rightarrow \gamma}=10^{-24}$ W, as expected in the range of models and of operating parameters for ADMX-HF,  corresponds to 0.3 photons/second at 5 GHz, or 30 photons in a 100 second averaging period.  This implies an intrinsic signal-to-noise of $\approx \sqrt{30}= 5.5$, or for the present example, a total signal-to-noise with thermal photons of $30/\sqrt{30+60}=3.2$.  This numerical example brings up an interesting issue: in the absence of thermal photons, the probability to observe at least one axion-generated photon over the observation time $t$ should be at the 95\% confidence limit. From Poisson statistics, the probability to observe zero events given a mean observation number of $\lambda$ is $P(0,\lambda)=\exp{(-\lambda)}$ which is $0.05$ for $\lambda\approx 3$. This can be taken as the minimum detectable axion signal for photon counting at zero temperature.  For a non-zero temperature, the minimum detectable power can be determined from the usual criterion that the signal-to-noise ratio be at least 5.  Thus, for the present numerical example, the signal-to-noise for a $10^{-24}$ W signal is only 3.2, which is slightly below minimum detection limit criterion.

Returning now to the principal issue of the photon counting shot noise fluctuations, assuming only thermal photons, Eq. (8) leads directly to an equivalent photon counting noise power,
\begin{linenomath}\begin{equation}
P_{sp}={h \nu \delta N\over t}= h\nu \sqrt{\eta \bar n \Gamma\over t}.
\end{equation}\end{linenomath}
Comparing the photon shot noise power and linear amplifier noise power,
\begin{linenomath}\begin{equation}
{P_\ell\over P_{sp}}={\bar n + 1 \over  \sqrt{ \bar n } }\sqrt{\Delta \nu_a\over \eta\Gamma}.
\end{equation}\end{linenomath}
This can be cast into a simpler form by noting that $\Gamma=2\pi \nu/Q_c$;  also, the optimum bandwidth for linear detection of the axion  signal is $\Delta \nu_a\approx  \nu/Q_a$. Therefore, in terms of the thermal photon occupation number, $Q_a$, and $Q_c$,
\begin{linenomath}\begin{equation}
{P_\ell\over P_{sp}}=\left[\sqrt{\bar n}+{1\over \sqrt{\bar n}}\right]\sqrt{Q_c\over \ 2\pi\eta  Q_a}.
\end{equation}\end{linenomath}
($\eta$ can be different for the two detection methods, and in general it is not the noise powers that need to be compared, but the $SNR$s.  Given that $\eta\sim 1$, the comparison of the noise powers is adequate for the present discussion.)

In the case of ADMX-HF, operating at a modest temperature of 100 mK, $\bar n=.1$, and conservatively $Q_a/Q_c\approx 20$ so (with $\eta\approx 1$)
\begin{linenomath}\begin{equation}
{P_\ell\over P_{sp}}\approx\left({1\over .3}+.3\right)\times {1\over 9.5}\approx {1\over 3}
\end{equation}\end{linenomath}
and it appears that linear detection is better than photon counting detection for these parameters.

On the other hand,  if we pick a readily achievable temperature $T
= 10$ mK, then $\bar n\approx 3\times 10^{-11}$ then $P_\ell/ P_{sp} \sim 17,000$ and clearly single photon detection has lower noise (and we need to include dark count rates and other relevant contributors to the noise assessment).
Similarly,  for a conservative temperature $T= 100$ mK, but with an improved cavity quality factor
$Q_a/Q_c= 3$,  $P_\ell/ P_{sp} \sim 0.5$ so linear detection will contribute lower noise. With both lower temperature and enhanced quality factor together, single-photon detection is obviously superior, $P_\ell /P_{sp} \sim 30,000$. In these limits careful attention needs to be paid to the dark count rate of single photon detection (see Summary below).

\section{Prospects for single-photon and squeezed-state detection}

We briefly discuss in turn the status of actual devices such as
Rydberg atom single-photon detectors and prospects of more conventional
quantum devices such as JPAs.  We are concerned here only with the fundamentals of the detection process and do not discuss the technical issues relating to their implementation in ADMX or ADMX-HF.

The possibility of single-photon detection in an ADMX-type experiment has already been demonstrated
in the CARRACK experiment, which utilized a Rydberg atom single-photon
detector \cite{5}. The temperature dependence of blackbody photons in the
cavity at 2.527 GHz, measured by the resonant absorption of the $111s_{
1/2}\rightarrow 111p_{3/2}$ transition in a $^{85}$Rb beam and
subsequent selective field ionization, was in perfect agreement with the
theoretical occupation number down to $T =$ 67 mK, a factor of
two below the SQL noise temperature. While sustained operation as an
axion search was not achieved at that time, no fundamental limitations
were encountered beyond the overall complexity of the technique.

Recent advances have simplified Rydberg atom microwave electrometry \cite{natphys}. These new methods have the electric field sensitivity required to detect the presence of a single photon in the ADMX-HF cavity, and are tunable over the 1-100 GHz range.  However, achieving this level of sensitivity requires a bandwidth of order 1 Hz.  Because a single photon will exist as a transient with the loaded cavity lifetime, the single photon detection bandwidth needs to be comparable to the cavity bandwidth, and this will require about a factor of 1000 improvement in signal to noise for these new methods, which might be possible.

JPAs, such as used presently in ADMX-HF, naturally generate squeezed
microwave fields and the technology of parametric amplification has
advanced sufficiently that it is realistic to imagine injecting a
squeezed field generated by one JPA into the axion cavity and then
detecting the field emerging from the cavity with a second JPA \cite{4a}.
The virtue of detecting a squeezed field with a JPA can be seen by writing the cavity's state as a quantum mechanical generalization
of a voltage phasor
\begin{linenomath}\begin{equation}
V_{cav}\propto\left[\hat q_0(t)\cos(\omega_c t)+\hat p_0(t)\sin(\omega_c t)\right]
\end{equation}\end{linenomath}
The two quadratures  $\hat q_0$ and $\hat p_0$ neither commute with the Hamiltonian nor with each other. Separately these non-commutations each contribute $h\nu/2$, and thus a total noise $h\nu$ to the system.
However, if the cavity is prepared not in its vacuum state, but in a squeezed state with reduced fluctuations in $\hat q_0$ (and consequently amplified fluctuations in $\hat p_0$), and one measures $\hat q_0$ only, the noise power could be suppressed to arbitrarily low values dependent on the degree of squeezing.
Presently, a receiver's system noise can be reduced to about $h\nu/4$, limited
by the power losses in microwave components and cables \cite{4}. Such a modest
improvement does not justify the additional complexity of the receiver
at this time, but with improvements in the efficiency of microwave components,
squeezed state devices may soon be an attractive option.


Superconducting circuits designed to acts as quantum bits (qubits) seem well-poised to act as microwave photodetectors in an axion experiment. Indeed, superconducting qubits have been been operated as detectors of single microwave photons analogous to optical photomultiplier tubes\cite{6}, but as yet the dead-time and bandwidth of these detectors are not well suited to the axion signal.  More promising would be to adapt the the systems that embed a superconducting qubit inside of a microwave cavity\cite{Wallraff2004} (so-called circuit quantum electrodynamics (cQED)) to the task of photodetection. These cQED systems have demonstrated an astounding ability to control and detect microwave fields inside the cQED cavity itself. Specifically, the presence of a single microwave photon inside a cavity is sufficient to shift the qubit transition frequency by many qubit linewidths. Consequently, one can test for the presence of a single photon by applying a microwave pulse--resonant with the qubit but not the cavity--that brings the qubit to its excited state if and only if the cavity contains a single photon. By subsequently measuring the state of the qubit, one has answered the question ``is there a photon in the cavity?'' Remarkably, the whole procedure can be accomplished faster than a photon decays from the cavity\cite{kirchmair2013}. As such, one can envision coupling the radiation field emerging from the axion cavity into the cQED cavity and interrogating the cQED cavity sufficiently rapidly to match the axion cavity bandwidth. While many design challenges must be overcome to incorporate a state-of-the-art cQED experiment into an axion search, the scheme would be technically homogenous and naturally cryogenic.

%

\section{Summary}

In conclusion, the axion signal has a narrow bandwidth compared to the cavity where the axions are converted to RF photons, and the relative narrowness is of advantage in a linear detection system because we need only consider the cavity noise over a small bandwidth, whereas single photon (or bolometric) detection experiences the full cavity noise bandwidth resulting in shot noise in the detected thermal photons. Photon counting techniques become favorable at sufficiently low temperatures (10 mK for ADMX-HF) and higher $Q_c$ RF cavities.  However one must bear in mind that increasing $Q_c>Q_a$ will result in a loss of axion signal, implying an upper limit to the improvement by increasing $Q_c$.

In comparing the noise performance of single-photon vs. linear detection, the parameters of interest are the ratio of the axion and cavity $Q$'s, and the number of thermal photons in the cavity; in the limit of $k_BT/ h\nu<<1$,
\begin{linenomath}\begin{equation}
\left[{P_\ell\over P_{sp}}\right] \approx \sqrt{{Q_c\over 2\pi Q_a}e^{h\nu/k_BT}}
\end{equation}\end{linenomath}
and when this number is large, cavity photon counting can have a
lower noise than linear detection. At conservative values of $T=100$ mK and $Q_a/Q_c=20$, the crossover point occurs at a
surprisingly low frequency ($\sim$10 GHz), not far above where
the current round of experiments will be running. The device technology
to support such a strategy may already be at hand.

One must bear in mind, however, that there remains a fundamental limit to the detectable axion photon power. As discussed, a minimum number of three photons that must be detected to give 95\% confidence for the presence of an axion signal.  Thus, the minimum power, given a frequency $\nu$ and observation time $t$, is (which are 5 GHz and 100 sec for ADMX-HF)
\begin{linenomath}\begin{equation}
P_{min}={3 h\nu \over t}= 10^{-25}\ {\rm W}\left[ {\nu\over {\rm 5\ GHz}}{ 100\ {\rm sec}\over t}\right].
\end{equation}\end{linenomath}
Furthermore, to achieve this limit, the dark count rate must be suppressed to below one count during $t$ at the 95\% confidence level.  Therefore to observe zero photons with 95\% probability,
$P(0,\lambda)=\exp(-\lambda)=0.95$ or $\lambda\approx 0.05=1/20$, so that
\begin{linenomath}\begin{equation}
n\Gamma t= {1\over 20} =\Gamma t e^{-h\nu/k_BT_{max}}.
\end{equation}\end{linenomath}
This implies the physical temperature requirement
\begin{linenomath}\begin{equation}
T<T_{max}= {h\nu\over k_B\log(20\Gamma t)}\approx 14.5\  {\rm mK}\ \left[{\nu\over 5\ {\rm GHz}}\right]
\end{equation}\end{linenomath}
using typical parameters for ADMX-HF ($Q_c\approx 40,000$, $\nu=5$ GHz, $t=100$ sec).  These are quite fundamental limits on the minimum detectable axion power and on the maximum physical temperature for single photon counting to be fully effective.

This work was supported under the auspices of the National Science
Foundation, under grants PHY-1067242, and PHY-1306729, and the auspices
of the U.S. Department of Energy by Lawrence Livermore National
Security, LLC, Lawrence Livermore National Laboratory under Contract
DE-AC52-07NA27344.


\end{document}